\documentclass[aps,twocolumn,showpacs,lengthcheck,preprintnumbers]{revtex4-1}
\usepackage{amsfonts}
\usepackage{amsmath}
\usepackage{amssymb}
\usepackage{graphicx}
\usepackage{fontenc}
\newcommand{\floor}[1]{\lfloor #1 \rfloor}

\begin{document}
\title{Unlocked-relative-phase states in arrays of Bose-Einstein condensates}
	
\author{Dengling Zhang}
\affiliation{
	School of Science, Xi’an University of Posts and Telecommunications, Xi’an, China
}
\author{Haibo Qiu}
\affiliation{
	School of Science, Xi’an University of Posts and Telecommunications, Xi’an, China
}
\author{A. Mu\~{n}oz Mateo}
\affiliation{Departamento de F\'isica, Universidad de La Laguna, La Laguna, Tenerife, Spain}


\begin{abstract}
Phase engineering techniques are used to control the dynamics of 
long-bosonic-Josephson-junction arrays built by linearly coupling Bose-Einstein condensates.
Just at the middle point of the underlying discrete energy band of the system, unlocked-relative-phase states are shown to be stationary along with the locked-relative-phase Bloch waves. In finite, experimentally-feasible systems, such states find ranges of dynamical stability that depend on the ratio of coupling to interaction energy. The same ratio determines different decay regimes, which include the recurrence of staggered-soliton trains in the condensates around Josephson loop currents at the junctions. These transient solitons are also found in their stationary configurations, which provide striped-density states by means of either dark-soliton or bright-soliton trains.
Additionally, the preparation of maximally out-of-phase (or 
splay) states is demonstrated to evolve into an oscillation of the uniform density of the condensates that keeps constant the total density of the system and robust against noise at low coupling.
\end{abstract}

\maketitle

\section{Introduction}
The arrays of Josephson junctions, either in superconducting or 
nonlinear-optical systems, have been very successful in the development of
technical applications.
Three types of time periodic states have been studied in series arrays of 
Josephson Junctions: in-phase states (or locked-phase junctions), splay states (with the phases of the junctions evenly distributed), and incoherence states (with nonuniform distribution of junction phases) \cite{Watanabe1994}. 
Equivalent states have also been found in the related
systems of globally coupled (discrete) Ginzburg-Landau equations \cite{Hakim1992}.
As far as we know, only the first type of such stationary states, having locked-phase junctions, have been explored within the scope of bosonic Josephson junctions made by arrays of coupled Bose-Einstein condensates (BECs). This situation may derive from the fact that these systems lack in general the global-coupling arrangement of junctions used in superconductors or optics. Instead, BECs are usually connected by the next-neighbor coupling established through the barriers of optical-lattice potentials operating in a tight binding regime \cite{Morsch2006}. Ultimately, both configurations, next-neighbor and global coupling, can be considered as limit cases of linear coupling with different spatial ranges \cite{Zou2009}.

The Josephson effect was soon realized in ultracold-gas experiments
\cite{Smerzi2003,Albiez2005,Levy2007}, addressing mainly phenomena associated with single and short Josephson junctions. Regarding extended junctions,
special theoretical attention has been paid to systems of two linearly coupled one-dimensional (1D) BECs, which configure a single long bosonic Josephson junction. Beyond the symmetric and anti-symmetric uniform states typical of the point-like junction in a double-well potential, many works have focused on the stationary nonlinear waves known as Josephson vortices
\cite{Kaurov2005,Kaurov2006,Brand2009,Qadir2012a,Montgomery2013,
Gallemi2016,Sophie2018}, which have been recently observed in experiments 
\cite{Schweigler2017}.

Concerning the study of junction arrays, up to date only the arrays of point-like Josephson junctions have been experimentally realized \cite{Cataliotti2001}. Theoretically,
particular features of the arrays of long Josephson junctions 
have been explored, ranging from the superfluid-insulator transition \cite{Cazalilla2006},
the motion of bright solitons  \cite{Blit2012}, the exotic phases in the 
presence of gauge fields \cite{Budich2017}, the 
stabilization of sets of localized dark solitons and Josephson vortices \cite{Baals2018}, or the generation of transverse Josephson vortices \cite{Gil2019}.

\begin{figure}[tb]
	\centering
	\includegraphics[width=0.9\linewidth]{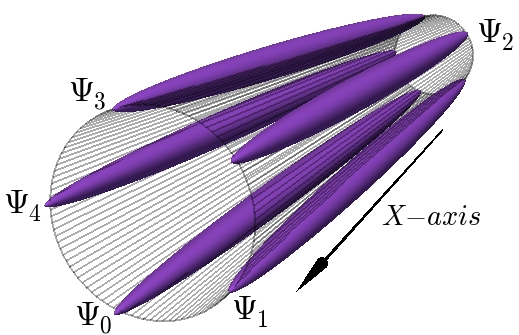}
	\caption{Array of five parallel, linearly coupled elongated BECs with order parameters 
		$\Psi_j(x,t)=\sqrt{n_j(x,t)}\exp{[i\Theta_j(x,t)]}$, and 
		$j=0,\,1,\,2,\,3,\,4$. The condensates form a ring configuration by means of
		a next-neighbor coupling (symbolically represented by the wired cylinder) of energy  $\hbar\,\Omega/2$ along their whole axial length. The coupling expands azimuthally a discrete transverse dimension $y$ of effective distance 
		$\delta y=\sqrt{\hbar/m\Omega}$ between condensates.}
	\label{fig:schema}
\end{figure}
In this work, we consider arrays of long-bosonic Josephson junctions that are brought about by the stack of linearly-coupled elongated BECs. The junctions are described through the
relative phases of next-neighbor condensates, and their dynamics is studied within the 
Gross-Pitaevskii theory. Our center of interest is the existence and stability of
array states whose junctions have no locked relative phases. On the one hand, we show that there exist a set of stationary states, living just at the middle of the 
discrete energy band, that, in spite of sharing energy and density profile with the Bloch 
waves, break their locked relative phase. Dynamically stable states of this type can be found at high coupling, hence they are relevant for experimental realization. Furthermore, we also find steady, stable configurations with striped density profiles and unlocked relative phases that break the translation symmetry of the equations of motion. 
On the other hand, we address the dynamics
of maximally out-of-phase states that in this way mimic the splay states in globally coupled junctions. Although stationary states of this type cannot be found in next-neighbor coupling arrangements, we show how these states evolve through oscillations of the uniform density of the condensates that preserve the total density constant against noise at low coupling. Different decay scenarios of unlocked-relative-phase states that involve the emergence of dark and bright solitons, and localized Josephson currents, demonstrating the interplay of transverse and axial dynamics, are discussed.

\section{Array of coupled elongated BECs}

Figure ~\ref{fig:schema} shows the  prototypical arrangement of the
considered arrays. In this example, the system is made of $M=5$ elongated BECs linearly 
coupled along their common axial $x$-direction, forming 
a ring-shaped array. A linear coupling of energy 
$\hbar\Omega/2$ connects next-neighbor components and determines an
effective transverse distance 
$\delta y =\sqrt{\hbar/m\Omega}$ between them. 
Along the axial direction, the interparticle interaction defines a healing length 
$\xi=\hbar/\sqrt{m\,g\,n}$, where $n$ is a characteristic 
axial atomic density of the BEC, $g>0$ is the contact-interaction 
strength, and $m$ is the atomic mass.
The ratio $\nu\equiv(\xi/\delta y)^2 =\hbar\Omega/gn$ regulates the
amount of particle tunneling across the condensate junctions. As it has been recently proposed 
\cite{Budich2017,Baals2018}, such a 
system is feasible to experimental realization with ultra-cold gases loaded in 
optical lattices. 

Within a mean field approximation at zero temperature, the dynamics of an 
$M$-condensate array follows the Gross-Pitaevskii (GP) equations
\begin{align}
i\hbar\frac{\partial\Psi_j}{\partial t}  =\left( 
\frac{-\hbar^2}{2m}\partial_{x}^2 + 
g \left\vert \Psi_j\right\vert ^{2}\right)\Psi_j -
\frac{\hbar\Omega}{2} \left( \Psi_{j-1}+\Psi_{j+1}\right) \,,
\label{eq:gp}
\end{align}
for the complex order parameters 
$\Psi_j(x,t)=\sqrt{n_j(x,t)}\exp{[i\Theta_j(x,t)]}$, 
with density $n_j(x,t)$ and phase $\Theta_j(x,t)$, of the elongated BECs 
$j=0,1,\dots,M-1$. The transverse dynamics inside each BEC is assumed to be 
frozen by means of a tight transverse confinement. For the sake of analytical treatment, we 
further assume translational invariance along the $x$-direction, hence no external potential enters Eq.~(\ref{eq:gp}). 
The Josephson dynamics in the elongated junctions separating the BECs 
will be described through the relative phases $\varphi_j(x,t)=\Theta_{j+1}(x,t)-\Theta_j(x,t)$ and relative densities $\varrho_j(x,t)=n_{j+1}(x,t)-n_j(x,t)$.

The periodic configuration along $y$ admits stationary states in the form 
of transverse Bloch waves with transverse quasimomentum $\hbar\mathcal{K}_k$,
\begin{align}
\Psi_{j,k}(x,t)=\psi(x)\,\exp{[i(\mathcal{K}_k y_j-\mu_{k} 
	t/\hbar)]} \,,
\label{eq:Bloch}
\end{align}
where $y_j=j\,\delta y$ represents the discrete transverse coordinate, and $\mu_k$ is the chemical potential. The quasimomentum can take only $M$ different integer values within the first Brillouin zone $\mathcal{K}_k={2\pi\,k}/{M \delta y}$ with 
$k\in\left\lbrace 0,\,\pm 1,\,\pm 2,\dots,\floor{{M}/{2}}\right\rbrace$,
where $\floor{{M}/{2}}$ is the greatest integer less than or equal to 
$M/2$. As a result $\mu_k$ takes values within a discrete energy band of width $2\hbar\Omega$,  limited by the minimum and maximum values of $k$.
 
The axial wave function $\psi(x)$ entering Eq. (\ref{eq:Bloch}) can be whatever stationary solution of Eqs. (\ref{eq:gp}) at zero coupling $\Omega=0$ \cite{Carr2000}. The simplest case is the state with
uniform density $n$ and definite axial momentum $\hbar \mathcal{K}_x$, as
$\Psi_{j,k}(x,t)=\sqrt{n}\,\exp{[i(\mathbf{k\cdot r}_j-\mu_\mathbf{k} 
t/\hbar)]}$, where $\mathbf{k\cdot r}_j=\mathcal{K}_x \,x+\mathcal{K}_k y_j$ are the time-independent phases, and $\mu_\mathbf{k} = g n + {\hbar^2 \mathcal{K}_x^2}/{2m}- \hbar\Omega \cos(\mathcal{K}_k \delta y)$. All the Bloch states with 
constant density are dynamically stable if $k\leq M/4$, irrespective of the 
coupling $\Omega$ \cite{Gil2019}. 

\begin{figure}[tb]
	\centering
	\vspace{.5cm}
	\includegraphics[width=\linewidth]{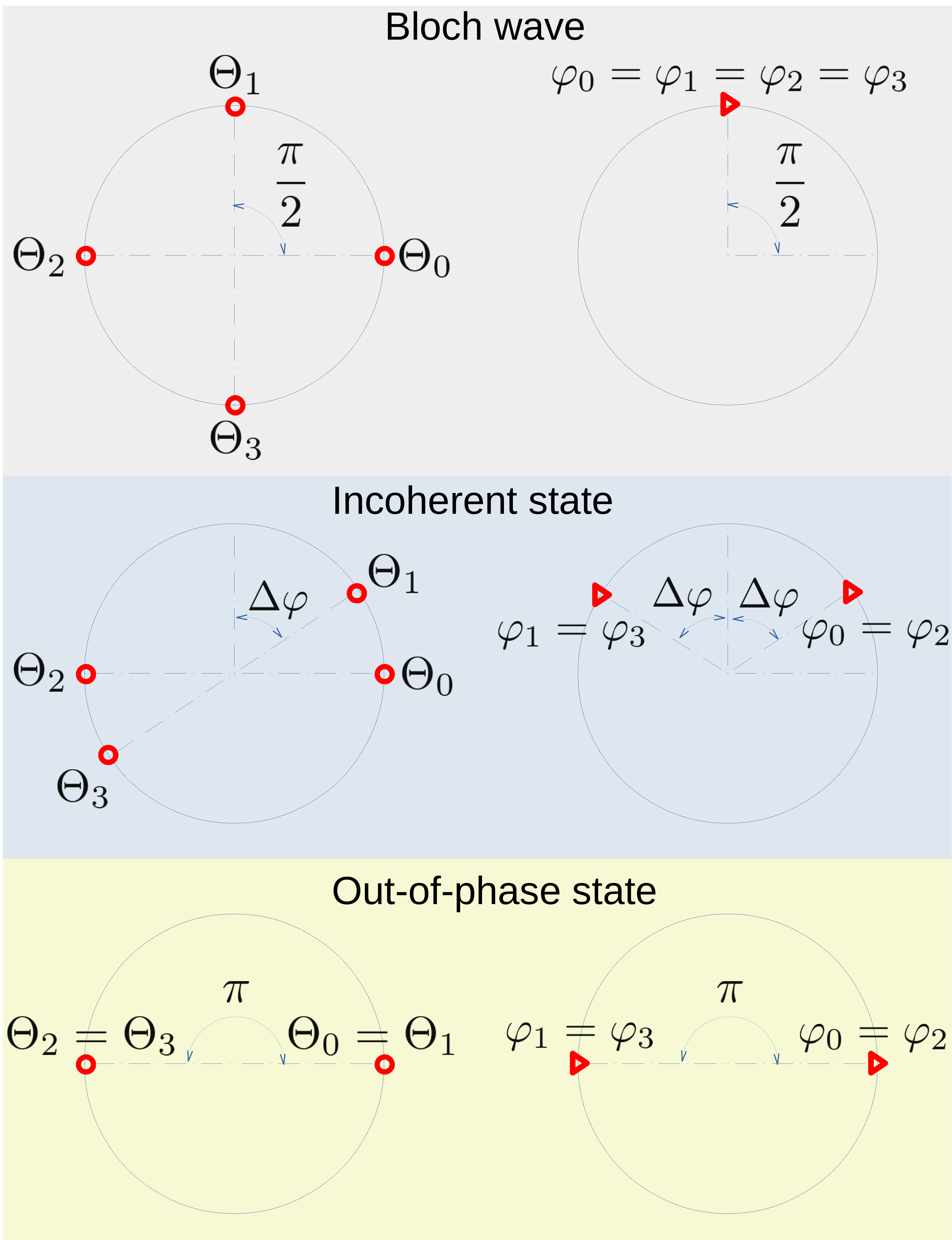}
	\caption{BEC phases $\Theta_j=\arg(\Psi_j)$ (circles) and relative phases 
		$\varphi_j=\Theta_{j+1}-\Theta_j$ (triangles), at $t=0$ and $x=0$, 
		of a $M=4$ array in a Bloch wave with $k=1$ (top panel), in an incoherent state (middle panel) and in an out-of-phase state (bottom panel), sharing the same chemical potential 
		$\mu_1=g n+\hbar^2\mathcal{K}_x^2/2m$, independently of the value of the linear coupling $\Omega$. }
	\label{fig:M4Bphase}
\end{figure}
In a generic Bloch state, for a given time and a fixed axial position, the phases of the BEC components of quasimomentum $\mathcal{K}_k$ are uniformly separated $\Theta_j(x,t)=2\pi k j/M+\Theta(x,t)$, so that the relative phases 
$\varphi_j=2\pi k /M$, $\forall\, j$, are locked everywhere and for every time. The top panel of Fig.~\ref{fig:M4Bphase} shows schematically this fact for a $M=4$ array in a dynamically stable, uniform Bloch wave with $k=1$, just at the middle of the discrete energy band.
The BEC phases of the array 
$\Theta_j(t=0)=[0,\frac{\pi}{2},\pi,\frac{3\pi}{2}]$ (circles in the left chart), and the relative phases $\varphi_j=[\frac{\pi}{2},\frac{\pi}{2},\frac{\pi}{2},\frac{\pi}{2}]$ (triangle in the right chart) are represented on the unit circle, in a phasor diagram.

Interestingly, along with the Bloch waves, there exist alternative stationary states sharing axial wave function, but breaking the monotonic variation of the phase across the array. As we show below, they are degenerate with the corresponding Bloch wave living just at the middle of the energy band. These states exist by virtue of the discrete nature of the system, since equivalent states sharing energy and density profile with the Bloch waves can not exist in continuous periodic potentials. The middle and bottom panels of Fig.~\ref{fig:M4Bphase} depict stationary configurations of the array phases belonging to this set. We will generically refer to the states of this set as unlocked-relative-phase states, in contrast to the locked relative phases of the Bloch waves.

\section{Locked- and unlocked-relative-phase states}

\subsection{Stationary states}
\label{sec:steady}

The Bloch waves situated just 
at the middle of the energy band, that is with $k=M/4$ (as the case shown in 
Fig. \ref{fig:M4Bphase}), present  $\frac{\pi}{2}$ relative phases that add to ${\pi}$ phase jumps between second-neighbor condensates. This fact
cancels the coupling dependence in the GP Eq.~(\ref{eq:gp}), since
$\Psi_{j+1}+\Psi_{j-1}=0$, and thus in the chemical potential, e.g. for uniform states $\mu_{M/4}  = g 
n+\hbar^2\mathcal{K}_x^2/2m$. The resulting configuration resembles features of 
splay states in globally coupled oscillators \cite{Hakim1992}, which in turn model 
Josephson-junction arrays \cite{Watanabe1994}. There, the splay states are 
characterized by oscillator phases that add to neutralize the coupling. To do so, the phases
are maximally out of phase, evenly distributed around the unit circle in a phasor diagram, and their existence is accompanied by a high degeneracy induced by the all-to-all coupling. Although in the setup considered here
the nearest-neighbor connection imposes a more restricted scenario, the neutralization of the coupling term in GP equation (\ref{eq:gp}) leads also to new degenerate stationary states.

In canceling the coupling, the condensate phases and the relative phases in a Bloch wave with $k=M/4$ fulfill
\begin{align}
\Theta_{j+1}(x)-\Theta_{j-1}(x)=\varphi_j+\varphi_{j-1}=\pi \,.
\end{align}
This precise configuration allows us to introduce an extra degree of freedom, a phase $\Delta \varphi$ that modifies neither the energy nor the chemical potential of the system when added to every second component, so that
\begin{equation}
\begin{aligned}
\varphi_j=\frac{\pi}{2}+\Delta \varphi,\\
\varphi_{j-1}=\frac{\pi}{2}-\Delta \varphi.
\end{aligned}
\label{eq:deltaphi}
\end{equation}
The arbitrary phase $\Delta \varphi \in [-\frac{\pi}{2},\frac{\pi}{2}]$ can be added to and subtracted 
from the relative phases of consecutive junctions to get a new, degenerate stationary state. 
In this way the relative phases of the array, which are locked for the Bloch wave, become unlocked without energy cost. This degeneracy reflects the symmetry of the short range coupling between condensates. Simultaneously, the Josephson current $\mathcal{J}_j= \Omega \sqrt{n_j n_{j+1}}\,\sin \varphi_j$ flowing between components $j$ and $j+1$, which measures the particle tunneling through the junctions, is reduced by a factor $\cos(\Delta \varphi)$ with respect to the phase-locked configuration. Note that the condensate phases resulting from Eq.~ (\ref{eq:deltaphi}) are in general not uniformly distributed in a phasor diagram, which corresponds to a diagram of the so-called incoherent states \cite{Watanabe1994}. 

The set of stationary states generated by the operation given in Eq.~ (\ref{eq:deltaphi}) can only be found in arrays whose number of components $M$ is a multiple of four, where the Bloch wave with $k=M/4$ exists.
The simplest state of this type appears in an system with $M=4$. For instance, from a uniform Bloch wave with $k=1$, new steady configurations can be chosen as
$\Theta_j(x)=\mathcal{K}_x\,x+[0,\frac{\pi}{2}-\Delta \varphi,\pi,\frac{3\pi}{2}-\Delta \varphi]$,
and the relative phases read
$\varphi_j=[\frac{\pi}{2}-\Delta 
\varphi,\frac{\pi}{2}+\Delta\varphi,\frac{\pi}{2}-\Delta 
\varphi,\frac{\pi}{2}+\Delta\varphi]$. In particular, for $\Delta\varphi=\pi/2$
one gets a stationary configuration with $\Theta_j(x)= \mathcal{K}_x\,x+[0,0,\pi,\pi]$ and 
$\varphi_j=[0,\pi,0,\pi]$, which is 
represented in the bottom panel of Fig. \ref{fig:M4Bphase} and that we will refer to hereafter as out-of-phase state. In this limit case the Josephson current in the array cancels. This is the maximally out of phase, stationary pattern achievable in the relative phase diagram of a system with nearest-neighbor coupling. For higher-$M$ arrays, the same phase pattern can be found repeated in out-of-phase stationary states that are degenerate with the Bloch wave of $k=M/4$, that is $\Theta_j(x)=\mathcal{K}_x\,x+[0,0,\pi,\pi,0,0,\pi,\pi,\dots]$, and 
$\varphi_j=[0,\pi,0,\pi,0,\pi,0,\pi,\dots]$. These phase patterns reflect the configurations associated with the boundary of the Brillouin zone in a supercell structure with double period. They can also be seen as the phases induced by a train of transverse dark solitons whose nodes are located at every other junction of the array.

 \begin{figure}[tb]
	\centering
	\includegraphics[width=\linewidth]{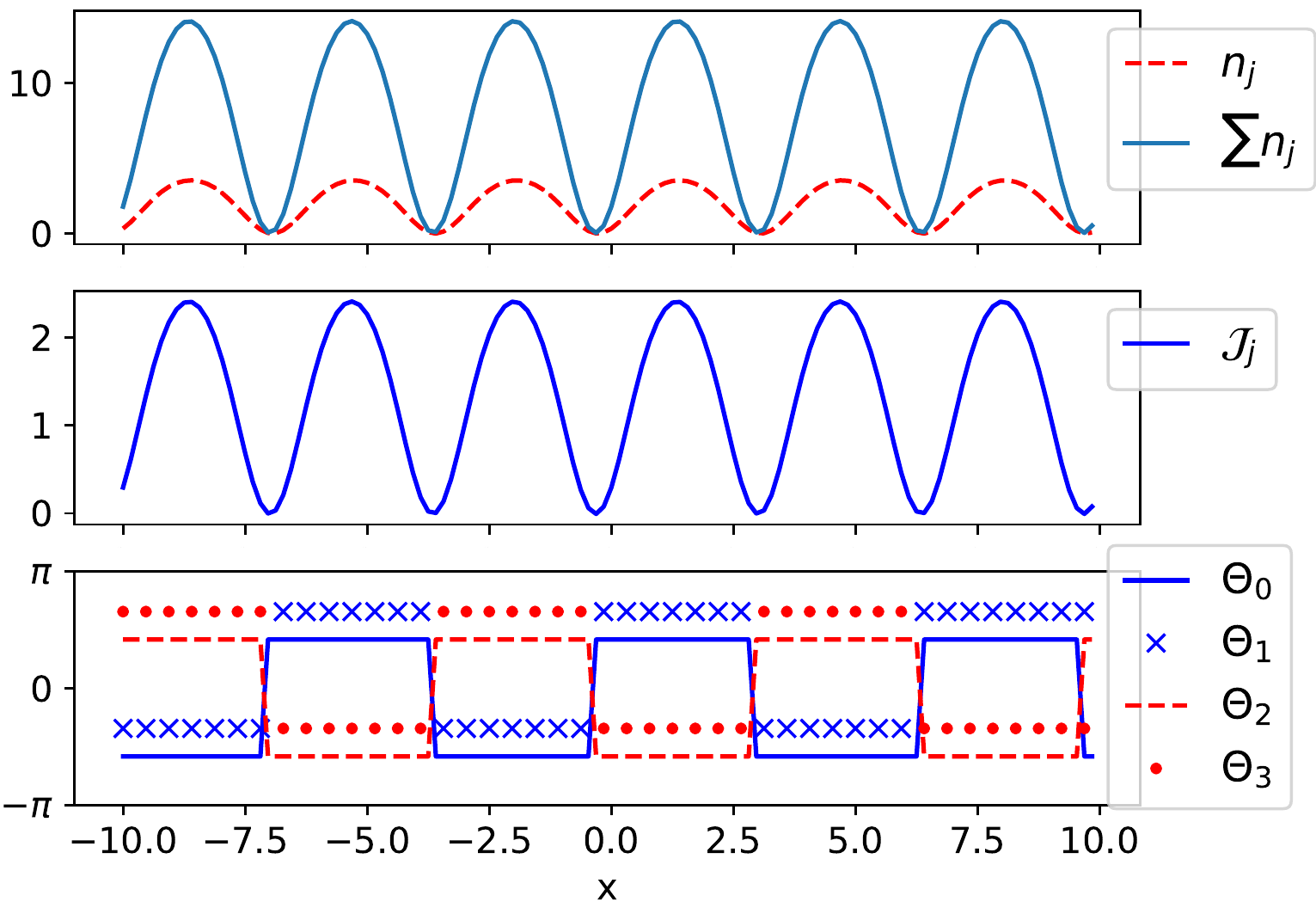}
	\caption{Unlocked-relative-phase soliton train in an array of $M=4$ condensates. The BEC densities $n_j$  (top panel) and the Josephson currents $\mathcal{J}_j$ (middle panel) are given in arbitrary units. As for the BEC phases (bottom panel), they produce the position-independent set of relative phases $\varphi_j=[\frac{\pi}{2}-\Delta 
		\varphi,\frac{\pi}{2}+\Delta\varphi,\frac{\pi}{2}-\Delta 
		\varphi,\frac{\pi}{2}+\Delta\varphi]$, with $\Delta\varphi=-\pi/4$. The chemical potential $\mu=2.64\,gn$ (see text) is independent of the coupling.  The length is measured in units of $0.45\,\xi$. }
	\label{fig:soliton3}
\end{figure}
The set of unlocked-relative-phase states includes also non-uniform configurations. As example, we present a family of states composed of trains of unlocked-relative-phase solitons that provide a periodic density profile or striped density. As in the uniform case, the soliton phases cancel the coupling terms in the equations of motion (\ref{eq:gp}), therefore the chemical potential is independent of $\Omega$. Figure \ref{fig:soliton3} depicts the steady configuration of these solitons in a finite system of axial length $L=12.5\,\xi$ (with periodic boundary conditions) and chemical potential $\mu=2.64 \,gn$, where we have defined an average density $n=N/L$ through the number of particles $N$ in each BEC. It corresponds to an analytical solution  of the sine-Jacobi type \cite{Kanamoto2009},
$\Psi_j(x)\propto \mathrm{sn}({12K} x/L,0.629) \exp{(i\Theta_j)}$ with elliptic modulus $\mathrm{m}=0.629$ and $K$ being the complete elliptic integral of the first kind. The soliton trains of all the BECs present overlapping density profiles, whereas the relative phases have been unlocked by an angle $\Delta\varphi=-\pi/4$.

\subsubsection{Linear stability of uniform states} 
The linear stability of unlocked-relative-phase states, with uniform density $n$ and generic stationary phases $\Theta_j(x)=\mathcal{K}_x\,x+[0,\frac{\pi}{2}-\Delta \varphi,\pi,\frac{3\pi}{2}-\Delta \varphi,\dots]$, can be found from the Bogoliubov equations 
for the linear excitations $[u_{j}(x), v_{j}(x)]$ of
energy $\mu+\hbar\,\omega$ \cite{Pitaevskii2003}:
\begin{equation}
\begin{aligned}
  H \, u_j+  \, g  \, n \, e^{i2\Theta_j } v_j 
-\frac{\hbar\Omega}{2} \left(u_{j-1}+u_{j+1}\right)
 & = \hbar \omega \, u_j \,,
\\
 -H \, v_j -  \, g  \, n \, e^{-i2\Theta_j} u_j 
+\frac{\hbar\Omega}{2}  \left(v_{j-1}+v_{j+1}\right) &= \hbar \omega \, v_j \,,
\end{aligned}
\label{eq:bog}
\end{equation}
where $H = -(\hbar^2/2m)\partial_x^2+  2 g n -\mu$, and
$\mu=\mu_{M/4}  = g n+\hbar^2\mathcal{K}_x^2/2m$. 

\begin{figure}[tb]
	\centering
	\includegraphics[width=\linewidth]{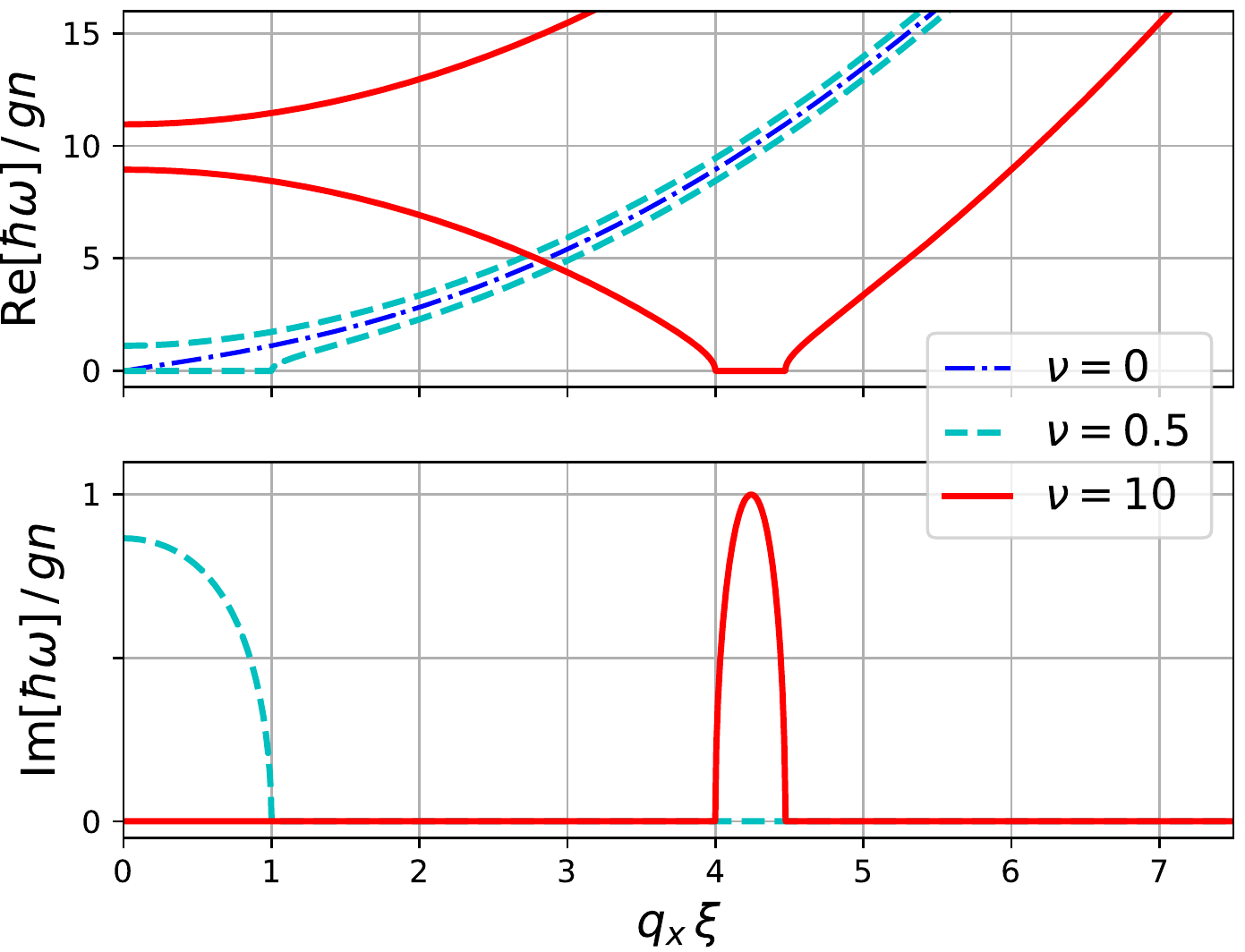}
	\caption{Dispersion of unstable modes of out-of-phase states for arrays of $M=4$ condensates in two different dynamical regimes determined by the ratio $\nu=\hbar\Omega/gn$. The curve labeled $\nu=0$, which provides the phonon spectrum at $q_x\rightarrow 0$, is common to both regimes. The high $\nu$ regime includes a branch with minimum-energy excitations, and also unstable modes, at nonzero axial momentum $q_x$, resembling features of a roton spectrum.}
	\label{fig:unstable}
\end{figure}
In the limit cases $\Delta \varphi=\pm\frac{\pi}{2}$, Eqs.~(\ref{eq:bog}) can be readily solved
by making use of the Fourier expansions 
$ u_j(x)=\sum_{\mathbf{q}} u_{\mathbf{q}}\exp\{i[ \mathcal{K}_x\,x +q_x\,x + q_p\, 
y_j]\}$ and 
$v_j(x)=\sum_{\mathbf{q}} v_{\mathbf{q}}\exp\{-i[ \mathcal{K}_x\,x -q_x\, x -q_p\, 
y_j]\}$, 
where $q_p=2\pi p/M\delta y$ is the transverse momentum of the excitation for 
integer $p=0,\,\pm 1,\,\pm 2,\dots \floor{M/2}$. The Bogoliubov equations get 
decoupled for each two-dimensional wave number $\mathbf{q}=(q_x,q_p)$, and the resulting dispersion is
\begin{align}
\label{eq:dispersion}
 \hbar \omega &= \frac{\hbar^2\mathcal{K}_x\,q_x}{m} + \\ 
 \pm &\sqrt{\left(\zeta_x-\hbar\Omega 
\cos\left(\frac{2\pi p}{M}\right)\right)\left(\zeta_x-\hbar\Omega 
\cos\left(\frac{2\pi p}{M}\right)+2gn\right)}, \nonumber
\end{align}
where $\zeta_x=\hbar^2q_x^2/2m$. The energy branches with $p<M/2$ produce imaginary frequencies $\omega$ that are associated with unstable modes.  
Due to their plane wave character, these modes are not localized. The maximum imaginary 
frequency leading the decay of the out-of-phase states depends on the ratio 
$\hbar\Omega\cos({2\pi p}/{M})/gn=\nu \cos({2\pi p}/{M})$. The analysis is simpler for $\mathcal{K}_x=0$. In this case, if such ratio is less than one, the maximum imaginary frequency has Im$[\hbar\omega]<gn$ and corresponds to $q_x=0$, otherwise it reaches Im$[\hbar\omega]=gn$ and corresponds to quasimomenta $q_x=\pm\sqrt{2(\nu\cos({2\pi p}/{M})-1)}/\xi$; for high coupling $\nu\gg 1$ the range of unstable modes becomes localized around the maxima (see Fig. \ref{fig:unstable}). As we show next, different decay dynamics result from each case, and the mentioned localization of the unstable modes in momentum space suggests a way to find stable steady configurations of out-of-phase states in finite systems, where the axial momentum can only take discrete values. Stability is found
when the set of these discrete momenta do not sample the small ranges of unstable modes.

An interesting feature of the spectrum of linear excitations at high $\nu$ is the presence of roton-like excitations. Figure \ref{fig:unstable} shows this fact in the spectrum of a uniform, out-of-relative-phase state in an array of $M=4$ condensates and $\nu=10$. Among the four energy branches, two of them are degenerated (labeled with $\nu=0$), and only one 
(corresponding to excitations with $p=0$) gives rise to unstable modes with pure imaginary frequencies. As previously mentioned, these unstable modes appear at nonzero axial wavenumber $q_x\neq 0$. The resulting curve of real-frequency excitations (upper panel of Fig. \ref{fig:unstable}) closely resembles the dispersion of more complex systems containing roton modes, where the roton minimum reaches the zero energy axis and the roton modes become unstable (see, e.g., the discussion in Ref. \cite{Chomaz2018} on a dipolar quantum gas). 
The roton instability produces exponentially-growing standing waves made of the combination of the roton modes with opposite quasimomenta. As we show later, these modes cause density modulations of the uniform configuration and recurrences of soliton trains, and also reflect the existence of stationary states with a striped-density profile. It is worth noting that this instability, determined by Eq. (\ref{eq:dispersion}), is characteristic of the simpler system of two-linearly-coupled elongated condensates forming a $\pi$ junction.

\begin{figure}[tb]
	\centering
	\includegraphics[width=\linewidth]{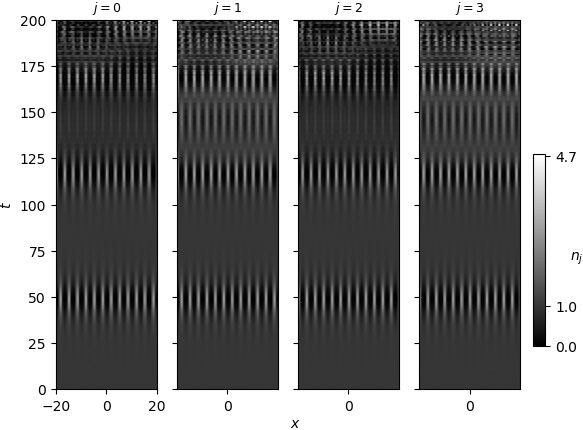}\\
	\vspace*{-0.5cm}
	\includegraphics[width=\linewidth]{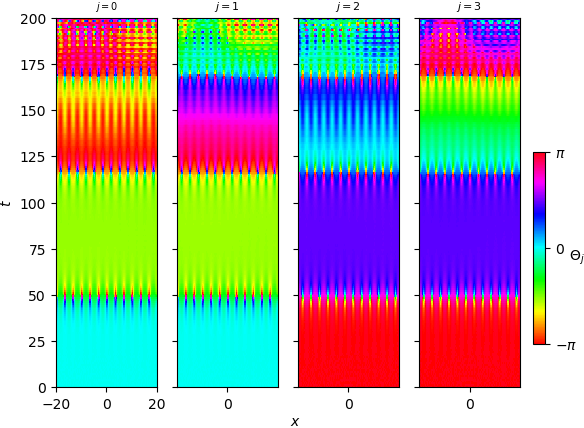}
	\caption{Time evolution of an out-of-phase state with $\nu=10$ in a array of  $M=4$ components. Component densities $n_j=|\Psi_j|^2$ (top panels) and phases $\Theta_j=\arg(\Psi_j)$ (bottom panels) are shown. The density is given in arbitrary units, whereas the time and length are measured in units of $5.0\,\hbar/gn$ and $0.45\,\xi$, respectively. }
	\label{fig:oof_nu10}
\end{figure}

\subsubsection{Dynamics} 
We report on the typical dynamics of out-of-phase states by numerically solving the GP Eqs.~\ref{eq:gp} for an array of $M=4$ condensates.
As has been shown in the linear analysis, the ratio $\nu=\hbar\Omega/gn$ determines the conditions for the stability of the system, and only the branch $p=0$ of Eq.~(\ref{eq:dispersion}) produces unstable modes. To demonstrate this fact, we have selected three case examples with $\nu=$ 0.5, 10, and 22, that represent respectively different dynamical regimes. In all of them, the system is constrained to evolve in an axially finite domain of length $L=17.9\, \xi$ and periodic boundary conditions. 
A white noise perturbation has been added to the initial stationary state in order to simulate a more realistic scenario.

The out-of-phase state is dynamically stable for $\nu=22$ since the unstable frequencies predicted by Eq.~(\ref{eq:dispersion}) occur for axial momenta (around $q_x=\pm6.48/\xi$) that are not sampled by the momenta $k=2\pi/L\times n=0.35/\xi\times n$, for integers $n$, determined by the finite system. For the given axial length, there are in fact many other instances of coupling above $\nu\approx 17.8$ that provide stability, e.g., systems with $\nu=19.9,\, 22,\, 24.5$ or 27 are equally stable. Our results of the nonlinear time evolution of the array in the presence of initial noise (not shown here because of their flat, uniform density and phase profiles) confirm the linear analysis.
In this way, the dynamical stability of these systems makes it possible their experimental realization.

\begin{figure}[t]
	\centering
	\includegraphics[width=\linewidth]{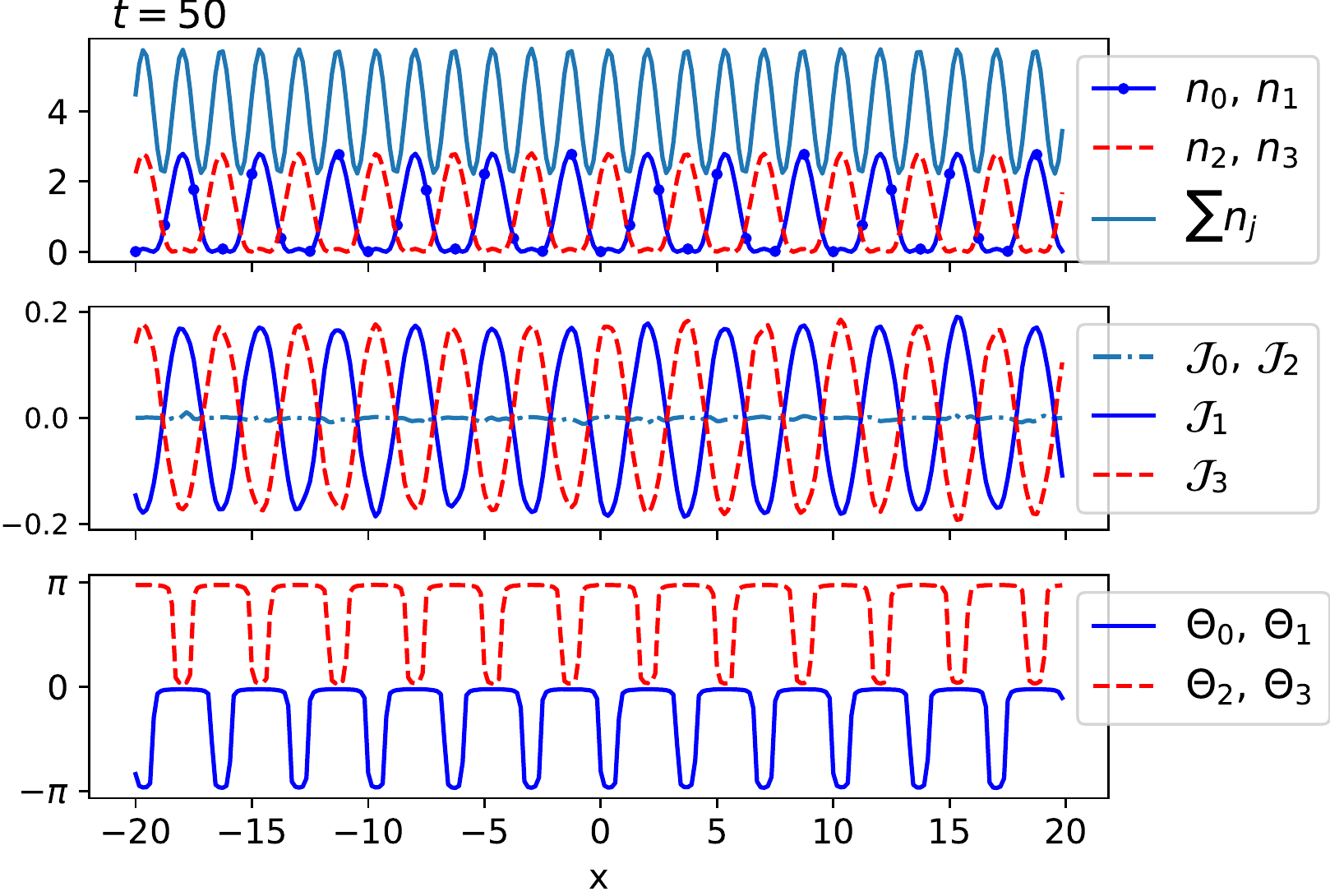}
	\caption{Transient staggered solitons during the emergence of the stripe phase in the evolution of the out-of-phase state presented in Fig. \ref{fig:oof_nu10}. At the junction between condensates $j=1$ and $j=2$, consecutive, counter-rotating loop currents are centered at the zeros of the Josephson current $\mathcal{J}_1$ and $\mathcal{J}_3$. The loops are closed by condensate axial currents, and their direction is reversed in the next soliton recurrence. Time, length and  densities are given in the same units of Fig. \ref{fig:oof_nu10}, whereas current is given in arbitrary units. }
	\label{fig:current}
\end{figure}
\begin{figure}[t]
	\centering
	\includegraphics[width=\linewidth]{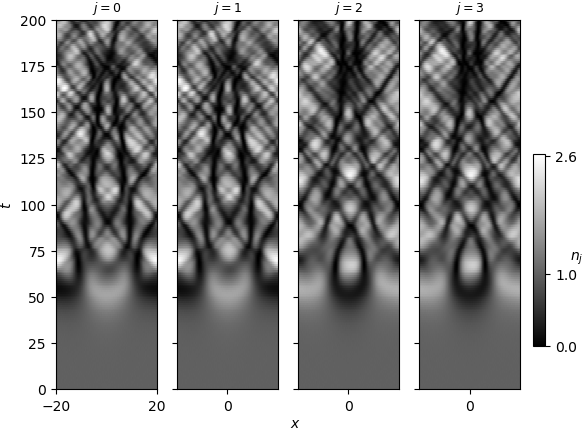}\\
	\vspace*{-0.5cm}
	\includegraphics[width=\linewidth]{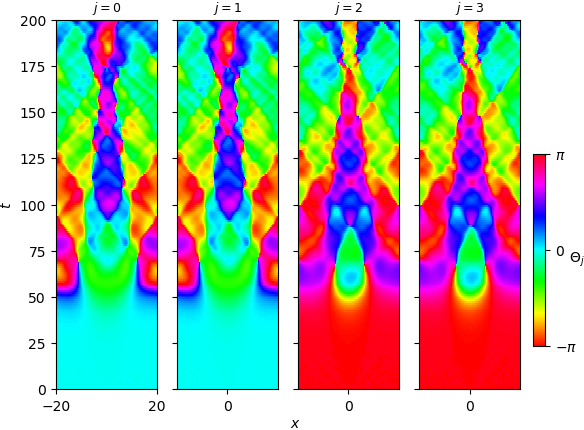}
	\caption{Same as Fig. \ref{fig:oof_nu10} for an out-of-phase state with $\nu=0.5$. }
	\label{fig:oof_nu05}
\end{figure}

 
However, for smaller coupling values, at $\nu=$10, and $\nu=$0.5, the instability cannot be prevented, and the out-of-phase configurations decay during the time evolution. 
An observable common feature, as can be seen in Figs. \ref{fig:oof_nu10} and \ref{fig:oof_nu05}, is the synchronous pattern shown by both densities and phases of different BECs. Interestingly, for initially in-phase components, they subsequently exhibit {\it in-phase} dynamics. The emergence of these synchronous patterns are due to the coupling in between BECs under highly symmetric arrangements of the arrays. The vanishing relative phases, apart from the perturbative noise, preclude significant Josephson currents that could break this synchronous pattern.

Nevertheless, the system dynamics presents notable differences in both cases of coupling. For $\nu=$10, Fig. \ref{fig:oof_nu10}  shows a quasi-periodic time evolution during which trains of solitons emerge in the axial direction of each component, breaking temporarily the uniformity of the $\pi$ relative phases and creating localized Josephson loop currents, and vanish, returning the array to its initial configuration. 
The number of solitons is given by the standing waves created with the wavenumbers  $q_x=\pm 4.2/\xi$ of the only two unstable frequencies (for this coupling strength). The solitons are staggered in the condensates with $\pi$ relative phase, which form trains of dark-bright solitons when the condensates are combined (see Fig. \ref{fig:current}). As a result, the total density profile of the system shows stripes of high contrast. The dark-bright sequence and also the Josephson loop currents are reversed with each new recurrence. The time recurrences of the solitons, thus of the striped state, suggest that this configuration is also dynamically unstable for the parameters considered here, as it also happens in similar types of modulation instability \cite{Kuznetsov2017}.

\begin{figure}[t]
	\centering
	\includegraphics[width=\linewidth]{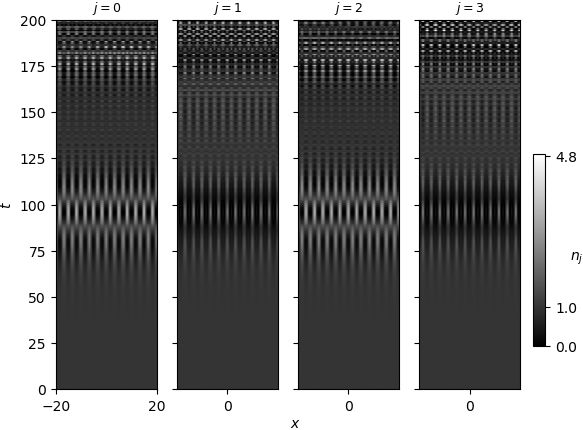}\\
	\vspace*{-0.5cm}
	\includegraphics[width=\linewidth]{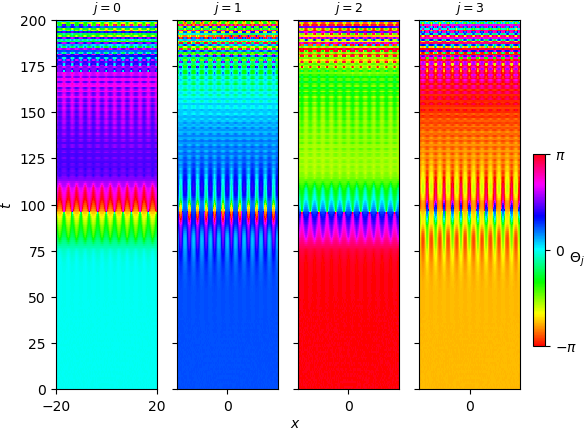}
	\caption{Same as Fig. \ref{fig:oof_nu10} for an unlocked-phase state with  $\Delta\varphi=\pi/4$ and $\nu=10$. }
	\label{fig:unlock_nu10}
\end{figure}
At low $\nu$ the decay of out-of-phase states  is characterized by the presence of several unstable modes whereby a quasiperiodic behavior cannot not reached. The interaction of unstable modes produces complicated scenarios that can soon give rise to chaotic dynamics. An example is shown in Fig. \ref{fig:oof_nu05} for $\nu=0.5$, where a few moving solitons and unsteady localized Josephson currents can be seen to emerge and interact within the array components. After this, the axial and transverse dynamics of the array are strongly coupled and the evolution increases progressively in complexity. 

The dynamical regimes of generic unlocked-phase states with $\Delta\varphi<|\pi/2|$ do not present significant differences with respect to those shown for out-of-phase states.  For detailed comparison, Fig. \ref{fig:unlock_nu10} shows the time evolution of an unlocked-relative-phase state with $\Delta\varphi=\pi/4$ and  $\nu=10$, sharing the rest of parameters with the  out-of-phase state of Fig. \ref{fig:oof_nu10}. In this case, the recurrences of the soliton trains present lower contrast and occurs at higher rate. 
Curiously, the synchronous pattern is only clearly
observable between second-neighbor condensates with initial $\pi$ relative phases. This is partly due to the fact that for these second-neighbor condensates, their equations of motion Eqs. (\ref{eq:gp}) share the same coupling terms.
 In addition, consecutive solitons in the soliton trains of component $j=0$ and $j=2$ evolve through merging, or alternatively splitting, in order to produce new, reversed staggered configurations   

\begin{figure}[t]
	\centering
	\includegraphics[width=\linewidth]{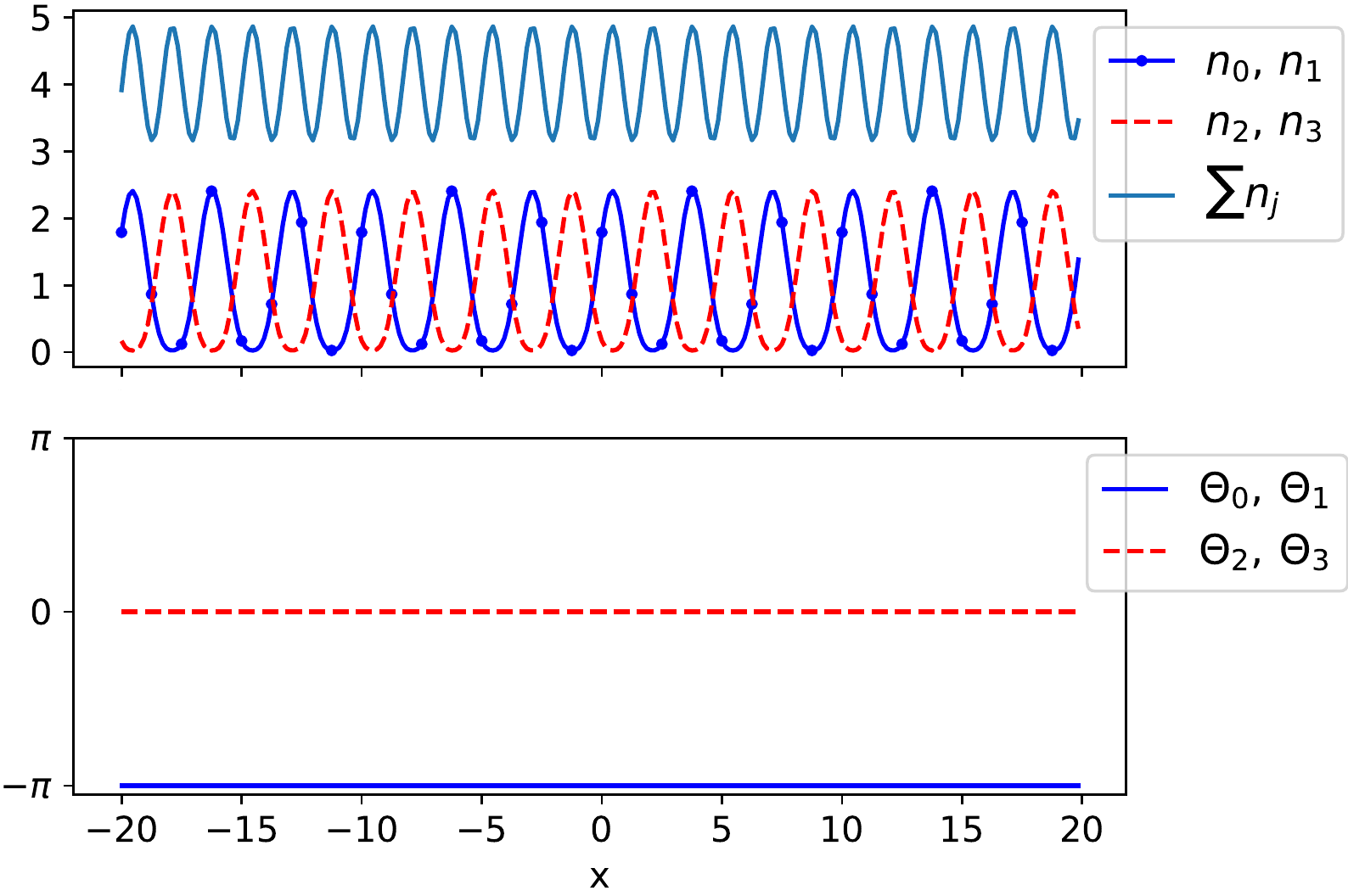}
	\caption{Stationary staggered solitons in an array of $M=4$ BECs with same parameters as the system of Fig. \ref{fig:oof_nu10}. Contrary to the transient configuration shown in Figs.  \ref{fig:oof_nu10} and \ref{fig:current} there is no Josephson current here.  }
	\label{fig:soliton12}
\end{figure} 
\begin{figure}[htb]
	\centering
	\includegraphics[width=0.85\linewidth]{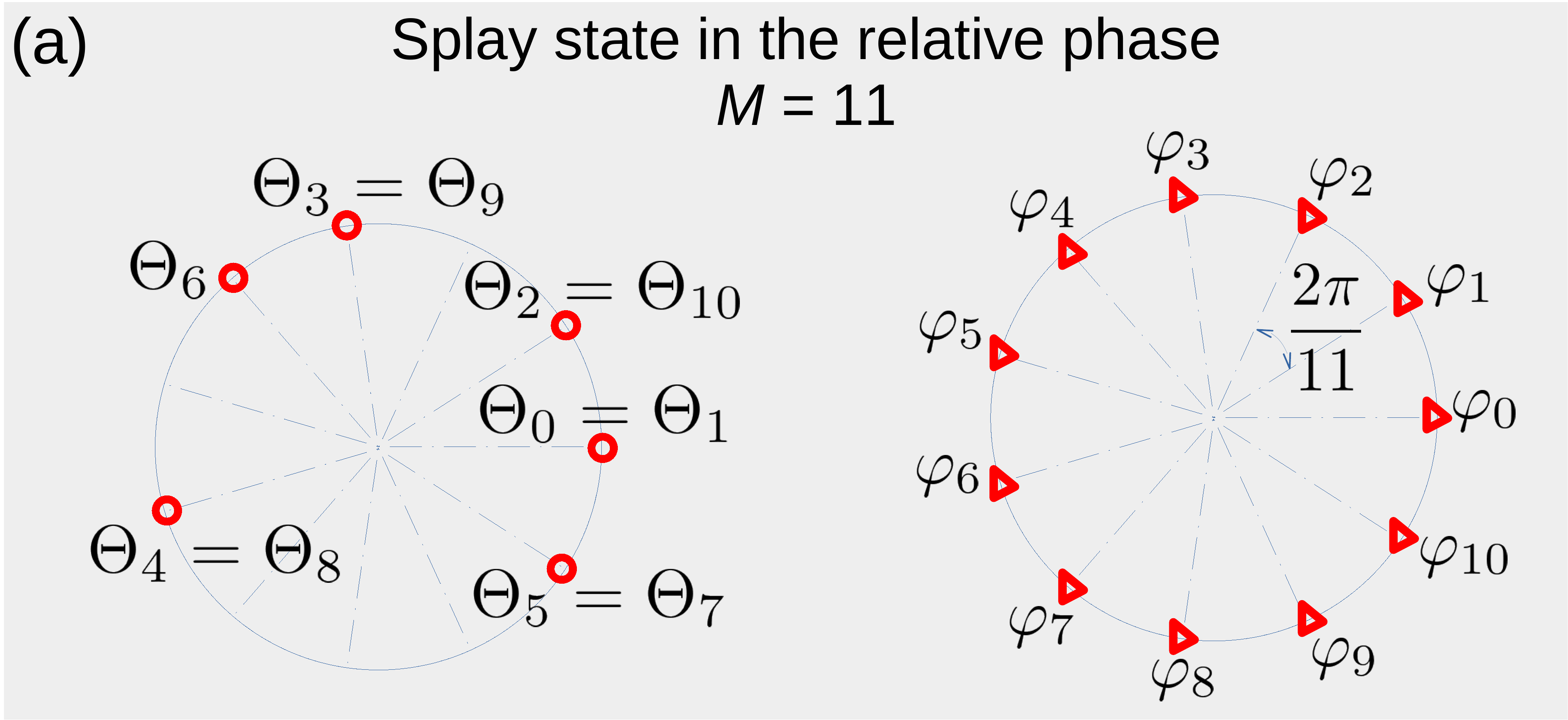}\\
	\includegraphics[width=0.85\linewidth]{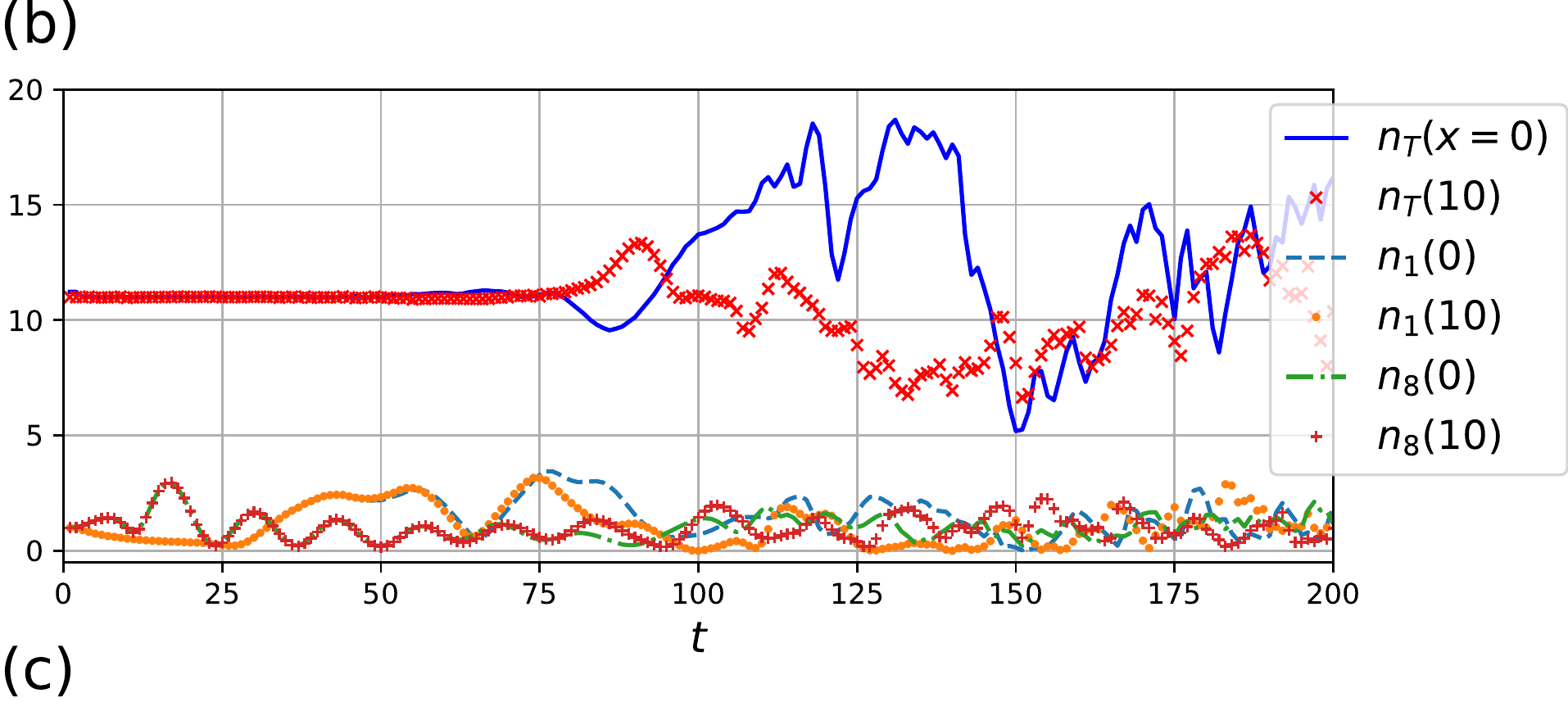}\\
	\includegraphics[width=0.9\linewidth]{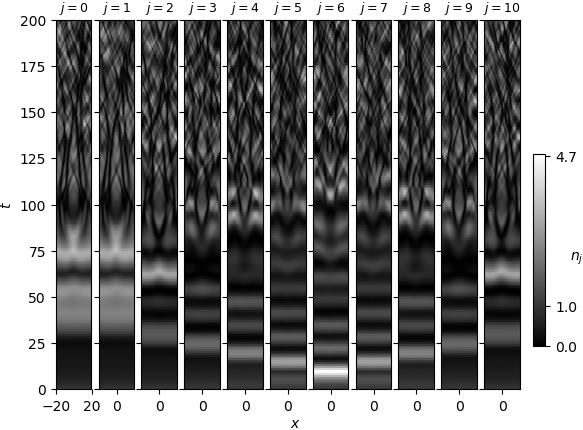}\\
	\vspace*{-0.5cm}
	\includegraphics[width=0.9\linewidth]{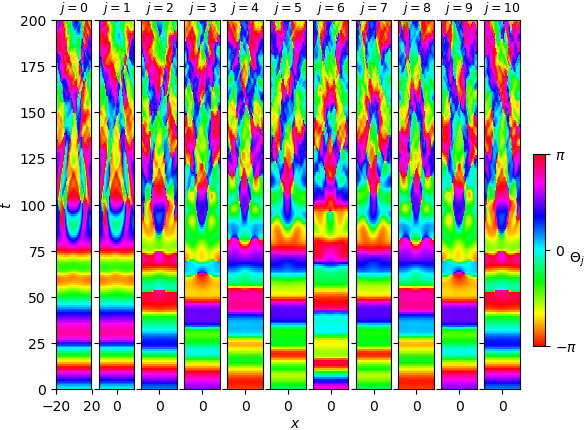}
	\caption{Maximally-out-of-phase state in an array with $M=11$ and $\nu=1$. (a) Initial BEC phases (left) and relative phases (right). (b) Time evolution of the axial (total $n_T$ and component $n_j$) densities at two positions $x=0,\, 10$. (c) Time evolution of the densities (top) and phases (bottom) of all the components. Same units as in Fig. \ref{fig:oof_nu10} are used.}
	\label{fig:splay_nu1}
\end{figure}
\subsubsection{Striped-density states: stationary staggered solitons}
The transient configurations that show striped density profiles at high coupling (see Figs. \ref{fig:oof_nu10} and \ref{fig:unlock_nu10}) suggest the existence of stationary configurations of this type. By seeding the uniform out-of-phase states with the unstable modes found in the Bogoliubov analysis, and by latter using a numerical Newton method to search for the solutions to Eqs.~(\ref{eq:gp}) with such ansatz, we have found the corresponding stationary striped-density states. They are made of out-of-phase bright solitons with staggered density profiles, in contrast to the overlapped-density dark-soliton trains presented in Sec. \ref{sec:steady}. Figure \ref{fig:soliton12} depicts an example for the same parameters and number of particles as the system evolved in Fig. \ref{fig:oof_nu10}. Note that there is no phase variation inside each condensate, and for this reason we have described the density profile as belonging to a bright-soliton train. This is a distinctive feature, since bright solitons do not appear in scalar BECs with repulsive interatomic interactions. Their appearance here can only be understood by the presence of the linear coupling arrangement, which, similarly to scalar condensates in periodic potentials, produces the change of sign in the effective mass of the particles \cite{Morsch2006}. The resulting scenario shows the emergence of a crystalline structure in the axial direction of the system without the presence of an external potential.

As expected, our numerical results show (by observing the decay in a real time evolution) that this configuration is dynamically unstable for the selected coupling. However, we have also found stable configurations of this type at high coupling, which brings these states into line with the stability conditions of uniform out-of-phase states.
The characteristic quantities of the stationary staggered solitons depend on the coupling. For given chemical potential, the contrast of the stripes decreases when the coupling increases, and eventually the uniform configuration of out-of-phase-states is reached. On the contrary, the density stripes get higher contrast by decreasing the coupling, which produces the generation of dark solitons. These structures finally transforms the staggered bright-soliton trains into a state of in-phase overlapped dark solitons, whose number doubles the initial number of staggered bright solitons in each condensate.

\subsection{Non-stationary maximally out-of-phase junctions}

For the sake of completeness, we also address the dynamics of array states whose relative phases are uniformly distributed on the phasor diagram (see Fig. \ref{fig:splay_nu1}a for an example with $M=11$). 
An stationary state of this type would be the analogue of the splay states in the array of globally coupled Josephson junctions \cite{Watanabe1994}. 
But differently to that situation, there is no such a stationary state in the array of next-neighbor-coupled BECs. 
Nevertheless, it is possible to statically prepare a splay state of relative phases in arrays with an odd number of BECs, and to use it as initial state in order to monitor its subsequent time evolution. To this end, one can choose among multiple settings of the BEC phases that produce static splay states in the relative phase. 
Although such an initial configuration generates unsteady Josephson currents in the array, the total density shows a quasi-stationary configuration in the absence of noise. 
This fact can be deduced from the hydrodynamical picture of Eqs. (\ref{eq:gp}), 
\begin{align}
\frac{\partial n_j}{\partial t}+\frac{\partial  n_j v_j}{\partial x}+\mathcal{J}_j-\mathcal{J}_{j-1}=0,
\label{eq:mass} \\
\hbar\frac{\partial \Theta_j}{\partial t}= \frac{\hbar^2}{2m\sqrt{n_j}}\frac{\partial^2 \sqrt{n_j}}{\partial x^2}-\frac{m v_j^2}{2}-g n_j + \frac{\mathcal{E}_j + \mathcal{E}_{j-1}}{2} \,,
\label{eq:momentum}
\end{align}
where $v_j=\hbar\partial_x \Theta_j/m$ and $\mathcal{E}_j=\hbar\Omega\sqrt{n_{j+1}/n_j}\cos\varphi_j$.
Then, whenever $v_j=0$ or $\partial_x \sum n_j v_j=0$, the total density $n_T=\sum n_j$ fulfills $\partial_t n_T=0$.
As a result, each array component presents an oscillatory density without axial variations while the total density is preserved. However, the presence of noise on the axial densities leads to axial currents that eventually produce the decay of the quasi-stationary configuration. 

As a case example, we present a maximally-out-of-phase state in an array with $M=11$ components and $\nu=1$.  The relative phases are evenly separated $\varphi_{j+1}-\varphi_j= \,2\pi/11$. The panels (b) and (c) of Fig. \ref{fig:splay_nu1} shows the subsequent time evolution after adding perturbative noise to the initial configuration. Within a first stage, up to $t\approx75$, the evolution follows the same described behavior as in the absence of noise. During this stage, nonlinear density waves, carried by Josephson currents that are uniform along the axial coordinate of each component, can be seen to propagate across the array. From inspection of Eq.(\ref{eq:mass}), one can see that the velocity of propagation is proportional to  $\sqrt{\hbar\Omega}$. Our results show that the duration of the quasi-stationary profile of the total density scales inversely with  this velocity.
In the present case, beyond $t\approx75$ the noise induces local variations of the Josephson currents between components that modify the flat density profiles. After this, the dynamics grows in complexity with time. It is worth noting that, despite the noise, the initially in-phase condensates keep synchronized density and phase profiles during a long time, of the order of the whole evolution.

\section{Conclusions}
The present work contributes to the characterization of the arrays of long-bosonic Josephson junctions built with linearly coupled Bose-Einstein condensates. In these systems,
we have demonstrated the existence and stability conditions of extended (non-localized) states with unlocked relative phases. These states emerge from the effective cancellation of the coupling in the equations of motion of the array, which allows for a new energy degeneracy in the system (other than the usual of the Bloch waves with equal absolute value of the quasimomentum) that is conditioned by the next-neighbor coupling of the discrete array.
Both uniform-density and dark-soliton-train states have been studied as prototypical examples.

 Regimes of stability, of quasiperiodic recurrence of striped density, and of complex, chaotic dynamics have been found depending on the ratio of coupling to interaction energy; and higher ratios favor stability. The typical decay dynamics of these states shares features with other modulation instabilities, proceeding through the condensate density variation according to the growth of standing waves created by unstable modes. Simultaneously, trains of counter-rotating Josephson loop currents centered at the junctions play equivalent role to that of regular vortex dipoles created at the nodal lines of dark solitons in continuous systems. 
 
 The Bogoliubov analysis of linear excitations have allowed us to determine the unstable modes of uniform states, and use them in the search of new stationary configurations with a striped-density profile.
 In doing so, and despite the respulsive character of the interatomic interactions considered, we have demonstrated the existence of staggered, stationary out-of-phase trains of bright solitons. By varying the linear coupling at fixed chemical potential, the family of these states evolves either through the generation of dark solitons (for decreasing coupling) or through the reduction of the density stripes (for increasing coupling). Consequently, it provides a bridge connecting the family of in-phase, overlapped dark solitons with the family of uniform out-of-phase states.

Finally, we have explored the preparation and dynamics of maximally out-of-phase states, which mimic the splay states in globally coupled junctions. Although the next-neighbor connection of the BECs do not allow for such stationary configurations, the prepared states evolve, in the absence of noise, in a quasi-stationary configuration that keeps constant the total density of the system, while the internal dynamics show large oscillations of the BEC densities. Our results show that low coupling regimes ($\nu\ll 1$) provide quasi-stationary states of this type which are robust against noise.

\bibliography{splay_bec}

\end{document}